\documentclass[12pt]{iopart}

\usepackage{graphicx}  
\usepackage{amssymb}  

\usepackage{xcolor}
\usepackage{ulem} 

\newcommand{\rh}{R_\mathrm{H}}
\newcommand{\ld}{\lambda_\mathrm{D}}
\newcommand{\kT}{k_\mathrm{B}T}
\newcommand{\eqref}[1]{(\ref{#1})}
\newcommand{\ec}[1]{\textcolor{magenta}{#1}}

\begin{document}

\title[DNA capture into the ClyA nanopore]{DNA capture into the ClyA nanopore: 
diffusion-limited versus reaction-limited processes}

\author{Stefanos K. Nomidis$^{1,2}$, Jef Hooyberghs$^2$, Giovanni
Maglia$^3$ and Enrico Carlon$^1$}

\address{$^1$ KU Leuven, Institute for Theoretical Physics,
Celestijnenlaan 200D, 3001 Leuven, Belgium}

\address{$^2$ Flemish Institute for Technological Research (VITO),
Boeretang 200, B-2400 Mol, Belgium}

\address{$^3$ Groningen Biomolecular Sciences \& Biotechnology Institute,
University of Groningen, Nijenborgh 7, 9747 AG Groningen, The Netherlands}


\begin{abstract}
The capture and translocation of biomolecules through nanometer-scale
pores are processes with a potential large number of applications, and
hence they have been intensively studied in the recent years. The aim
of this paper is to review existing models of the capture process by a
nanopore, together with some recent experimental data of short single-
and double-stranded DNA captured by Cytolysin~A (ClyA) nanopore. ClyA
is a transmembrane protein of bacterial origin which has been recently
engineered through site-specific mutations, to allow the translocation
of double- and single-stranded DNA. A comparison between theoretical
estimations and experiments suggests that for both cases the capture is
a reaction-limited process. This is corroborated by the observed salt
dependence of the capture rate, which we find to be in quantitative
agreement with the theoretical predictions.
\end{abstract}

%
%
%
%
%


\section{Introduction}

Current nanopore technologies offer a large number of interesting
applications for the analysis of DNA, proteins, peptides and other types
of small molecules~\cite{dekk07,venk11,sosk12,sosk15,huan17}. Such
devices detect the presence of single molecules by measuring a drop
in the ionic current passing through the pore. Two different types of
nanopores are presently used; solid-state nanopores can be fabricated by
various techniques that produce a small hole in a silicon~\cite{dekk07}
or graphene membrane~\cite{heer16}. The size and shape of these nanopores
can be tuned during the fabrication process.  Biological nanopores,
on the other hand, are proteins, typically of bacterial origin,
embedded within a lipid bilayer~\cite{venk11,shi16}. Compared to
solid-state nanopores the size of biological pore proteins cannot be
tuned, but they can be engineered with atomic precision by site-specific
mutations~\cite{ayub16,sosk13}. The most studied biological nanopore
is the alpha-hemolysin ($\alpha\rm{HL}$) protein, which is used in
the first commercial nanopore DNA sequencer~\cite{bayl15}.  Owing to
the narrow inner-pore constriction ($1.3$~nm), translocation through
$\alpha\rm{HL}$ is restricted to single-stranded DNA (ssDNA). While
nanopore DNA sequencers are based on the translocation of ssDNA, for
other applications it is desirable to consider pores also allowing the
translocation of double-stranded DNA (dsDNA). A recent review about
biological nanopore sensing and a discussion of commonly-utilized
nanopores can be found in Ref.~\cite{shi16}.

\begin{figure}[t]
\centering\includegraphics{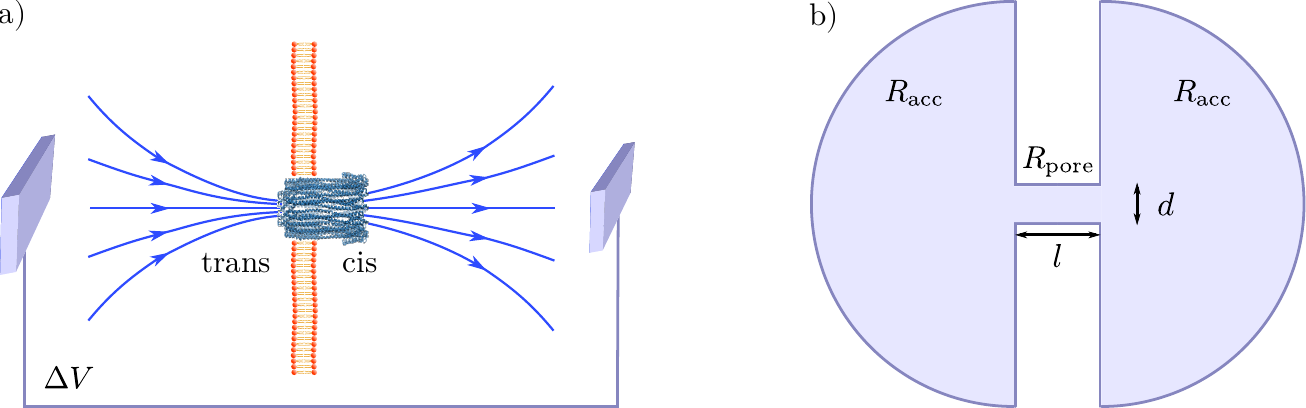}
\caption{a) Sketch of a typical experimental setup. A potential
difference $\Delta V$ is applied between two electrodes placed at
the two far sides of the membrane. The arrows indicate the electric
field lines direction. b) Physically, the system can be viewed as
a collection of three resistors in series, one corresponding to the
nanopore $R_\mathrm{pore}$, and two access resistances $R_\mathrm{acc}$
at its sides (shaded areas). Note that the latter are assumed to extend
to infinity.}
\label{fig:nanopore} 
\end{figure}

In this paper we analyze the capture of both ssDNA and
dsDNA by Cytolysin~A (ClyA), a biological nanopore which
has been recently employed both for nucleic acid and protein
analysis~\cite{sosk12,sosk13,sosk15,fran16}. In experiments, DNA
molecules are initially placed in the cis-side of the membrane.
An electric field is induced by applying a potential difference between
two electrodes placed at the two opposite sides of the membrane (see
Fig.~\ref{fig:nanopore}a). As a result, negatively-charged DNA molecules
diffusing in the vicinity of the nanopore are attracted to the pore entry.
After their eventual capture they either translocate to the trans-side,
or are released back to the cis-side.  Here we review the theory of the
DNA capture and discuss two possible mechanisms of diffusion-limited
and reaction-limited capture~\cite{gros10,rowg13}. We compare the two
mechanisms with experiments for short ssDNA and dsDNA captured by a
ClyA nanopore. We show that the dependence of these rates on the ionic
strength of the solution suggests that for both molecules the capture
is a reaction-limited process.

\section{The ClyA nanopore}

\begin{figure}[t]
\centering\includegraphics{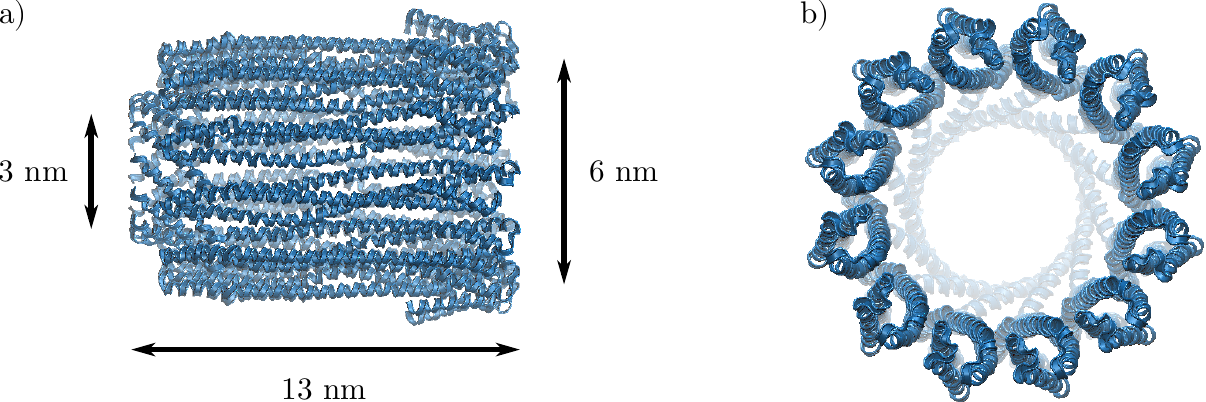}
\caption{Side (a) and top (b) view of the ClyA nanopore (PDB: 2WCD), with 
the characteristic size of the narrow wide and wide entries shown. For the 
visualization of the structure we used the VMD software~\cite{hump96}.}
\label{fig:clya} 
\end{figure}

Cytolysin A (ClyA) is a toxin synthesized by several bacteria, and is
employed to disrupt cellular membranes of other organisms. It is initially
synthesized as a monomer, and then it spontaneously assembles into a
12-mer, cylindrically-shaped pore (Fig.~\ref{fig:clya}). The internal
diameter is about $3$~nm at the narrower side and $6$~nm at the wider
side, while its length is $13$~nm.  Although the diameter of ClyA can,
in principle, fit both ssDNA and dsDNA, owing to the negative charges
present in the pore lumen, translocation in the wild type ClyA can only
occur in solutions of high ionic strength. For this reason ClyA mutants
were recently engineered~\cite{fran16}, that contain additional positive
charges in the lumen and the wide entrance of the pore, allowing DNA
translocation at physiological salt concentrations ($150$~mM NaCl).

\section{Modeling the DNA capture}

Figure~\ref{fig:nanopore}a shows a typical experimental setup, in
which a single nanopore is inserted in the lipid bilayer membrane, in
contact with a NaCl solution. When a potential difference $\Delta V$ is
applied between two electrodes, and in absence of blockages at the pore,
a steady electric current pore $I$ is generated, with oppositely-charged
ions flowing through the pore in opposite directions. The current can
be calculated by first decomposing the system into two semi-infinite
spherical shells (cis and trans side), connected with each other through
a cylinder of diameter $d$ and length $l$ (nanopore).  Then, treating
the three regions as resistors in series (Fig.~\ref{fig:nanopore}b)
and using Ohm's law yields~\cite{gros10}
\begin{equation} 
\Delta V = I \left(R_{\rm pore} + 2 R_{\rm acc} \right) =
\label{eq:DeltaV}
\frac{I}{\sigma} \left(\frac{4 l}{\pi d^2 } + \frac{1}{d} \right),
\end{equation} 
where $\sigma$ is the conductivity. We have denoted by $R_\mathrm{acc}$
the resistance of each semi-infinite half sphere, known as the access
resistance (the derivation can be found in Ref.~\cite{hall75}),
and by $R_\mathrm{pore}$ the electric resistance of the pore
(Fig.~\ref{fig:nanopore}b). The contribution of $R_\mathrm{acc}$ becomes
dominant for wider solid-state nanopores, as confirmed by experiments
with nanopores of varying $d$~\cite{kowa11}. In the case of ClyA,
which has dimensions $d = 6$~nm and $l = 13$~nm, one finds $2R_{\rm
acc}/R_{\rm pore} \approx 0.36$.  Assuming that the equipotential surfaces
are semi-spherical outside the pore, one obtains the electrostatic
potential~\cite{gros10}
\begin{equation}
V(r) =  \frac{I}{2\pi \sigma r} = \frac{\Delta V}{2 \pi r} 
\left(\frac{4 l}{\pi d^2 } + \frac{1}{d} \right)^{-1}
\equiv \Delta V \frac{\widetilde{d}}{r},
\label{Seq:cisV}
\end{equation} 
where we set the potential to zero at the electrode $r \to \infty$
and defined the characteristic length
\begin{equation}
\widetilde{d} \equiv \frac{1}{2\pi} 
\left(\frac{4 l}{\pi d^2 } + \frac{1}{d} \right)^{-1},
\label{Seq:qeff}
\end{equation} 
which depends only on the geometry of the pore (for the case of
ClyA one finds $\widetilde{d}=0.25$~nm). The DNA molecule performs a
drift-diffusive motion in the potential~(\ref{Seq:cisV}) until it reaches
the close vicinity of the pore. There it is either directly translocated
to the other side of the membrane, corresponding to a diffusion-limited
case, or it encounters an additional free energy barrier that needs
to overcome for a successful translocation. If the barrier is large
compared to the thermal energy $\kT$ and the attractive electrostatic
potential, it will dominate the capture kinetics, and the process becomes
reaction-limited. We will discuss these two cases separately, following
closely the theory developed in~\cite{gros10}.

\subsection{Diffusion-limited capture}

In spite of its high complexity, far from the pore the problem becomes
spherically symmetric [see Eq.~\eqref{Seq:qeff}], and DNA can be
treated as a charged point particle. Let us consider a collection of
such diffusing particles characterized by a concentration $c(\vec r,t)$
and subject to an electrophoretic force, given by a radial potential
$V(r)$. The continuity equation in spherical coordinates reads
\begin{equation}
\frac{\partial c}{\partial t} = 
-\frac{1}{r^2} \frac{\partial}{\partial r} \left[r^2 j(r) \right],
\label{Seq:diff}
\end{equation}
where the radial current density contains the contribution from diffusion
and electrophoretic drift
\begin{equation}
j(r) = - D \frac{\partial c}{\partial r}
+ \mu c \frac{\partial V}{\partial r}.
\label{jr}
\end{equation}
Combining Eqs.~\eqref{Seq:diff} and~\eqref{jr} one obtains the
drift-diffusion equation, with $D$ and $\mu$ the diffusion coefficient
and the electric mobility, respectively. Note that the two terms in
Eq.~\eqref{jr} enter with a different sign because DNA is negatively
charged, and by convention $\mu > 0$. It should be stressed that
the Einstein relation does not hold for free electrophoresis of
DNA~\cite{nkod01}, i.e.\ $D \neq \mu \kT$. This arises from the fact that
a free DNA in solution is accompanied by a collection of counterions,
while the application of an electric field pushes the two in opposite
directions.  This leads to different typical molecular configurations,
hence the Einstein relation breaks down.

The stationary solution of Eq.~(\ref{Seq:diff}) is obtained by setting
$\partial c/\partial t =0$, corresponding to constant $r^2 j(r)$. For
the $1/r$ potential of Eq.~\eqref{Seq:qeff} one finds~\cite{gros10}
\begin{equation}
c(r) = c_0 
\, \frac{1-e^{-r^*(1/R - 1/r)}}{1-e^{-r^*/R}},
\label{Seq:cr}
\end{equation}
where $c_0$ is the bulk concentration and $r^*$ a characteristic length
given by
\begin{equation}
r^* \equiv \frac{\mu \widetilde{d} \Delta V}{D}.
\label{Seq:defr*}
\end{equation}
For the derivation of Eq.~\eqref{Seq:cr} we used as boundary conditions
$\lim_{r\to\infty} c(r) = c_0$ and $c(R) =0$, with $R$ a microscopic
distance of the order of the pore size. From Eq.~(\ref{Seq:cr}) one can
estimate the capture rate, which is equal to the number of particles
per unit time reaching the absorbing boundary at $r=R$. This is obtained
by integrating the current density on a half-spherical shell of radius
$r$~\cite{gros10}
\begin{equation}
k_\mathrm{on} = 2 \pi r^2 j(r) = \frac{2 \pi D r^* c_0}{1-e^{-r^*/R}}  
\approx  2 \pi D r^* c_0,
\label{Seq:kon}
\end{equation}
where we have used $r^* \gg R$, which is a valid approximation for typical
systems~\cite{fran16}. Here $r^*$ can be interpreted as the 
distance at which the DNA is irreversibly captured by the pore~\cite{gros10}, 
and increases with the applied potential and the electrophoretic mobility [see 
Eq.~\eqref{Seq:defr*}]. Equation~\eqref{Seq:kon} is identical to the 
Smoluchowski diffusion-limited reaction rate for a free diffusing particle 
absorbed by a sphere of radius $r^*$, with $2\pi$ instead of $4\pi$ due to the 
semi-infinite geometry~\cite{gros10}.

\subsection{Reaction-limited capture}

In a reaction-limited capture the actual translocation takes place once
DNA overcomes an additional barrier at the pore entry. Ref.~\cite{rowg13}
discussed this type of process, which we review here. Let us consider a
drift-diffusion model with an additional short-range repulsive potential
$U(r)$, i.e.\ nonvanishing only in the close vicinity of the pore. The
radial current density is then given by
\begin{equation}
j(r) = - D \frac{\partial c}{\partial r}
+ \mu c \frac{\partial V}{\partial r} 
- \widetilde{\mu} c \frac{\partial U}{\partial r},
\end{equation}
where we distinguish between the electrophoretic mobility $\mu$
and a mobility $\widetilde{\mu}$ connected to other external
forces~\cite{long96}. Although the former does not satisfy the Einstein
relation, the latter does~\cite{rowg13b}, i.e.\ $D = \widetilde \mu \kT$.
Thus, we one can rewrite the current as
\begin{equation}
j(r) = - D \left( \frac{\partial c}{\partial r}
+ \frac{c}{\kT} \frac{\partial \Phi(r)}{\partial r} 
\right),
\label{Seq:jr_eff}
\end{equation}
with
\begin{equation}
\Phi(r) \equiv  U(r) - \frac{\mu}{\widetilde\mu} V(r).
\label{Seq:Veff}
\end{equation}

\begin{figure}[t]
\centering\includegraphics[width=8cm]{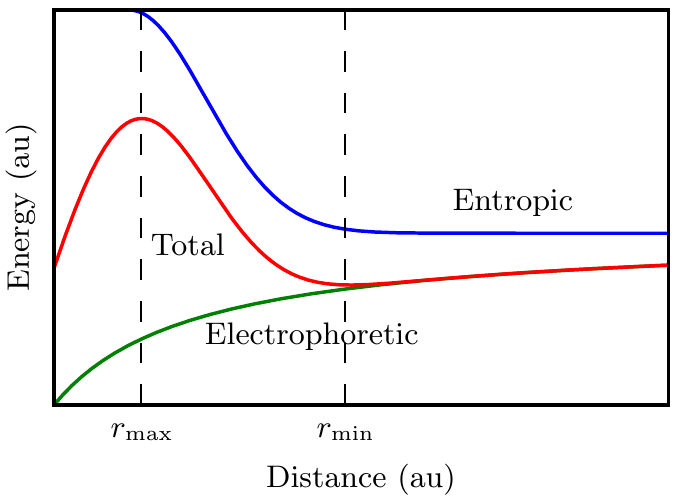}
\caption{Schematic plot of the effective potential in the vicinity of
the pore for a reaction-limited capture. In addition to the contribution
of the long-ranged, attractive electrophoretic force, a substantial
short-ranged repulsive force is expected to arise due to steric effects
for DNA threading into the pore~\cite{gros10,muth10}. The sum of the two
potentials has a minimum in $r_{\rm min}$ and a maximum in $r_{\rm max}$.}
\label{fig:energy}
\end{figure}

Thus, the dynamics is described by a drift-diffusion equation in the
effective potential $\Phi(r)$.  The electrophoretic contribution to the
force is attractive, while $U(r)$ is short-range and repulsive and
expected to originate from the steric hindrance of the DNA threading
into the pore~\cite{gros10,muth10} (Fig.~\ref{fig:energy}). To
initiate translocation the molecules have to overcome a barrier
$\Delta \Phi = \Phi(r_{\rm max}) - \Phi(r_{\rm min})$, where
$r_{\rm max}$ and $r_{\rm min}$ are the positions of the maximum and
minimum of $\Phi$, respectively, located in the vicinity of the pore
(Fig.~\ref{fig:energy}). If the capture process is reaction-limited
(i.e.\ $\Delta\Phi \gg \kT$), we expect that the rate will be given by

\begin{equation} 
k_\mathrm{on} = \omega \ e^{-\Delta \Phi/\kT},
\label{Seq:kon_ssDNA} 
\end{equation} 
where $\omega$ is a characteristic rate constant.

\section{Experiments}

\begin{figure}[t]
\centering\includegraphics[width=8cm]{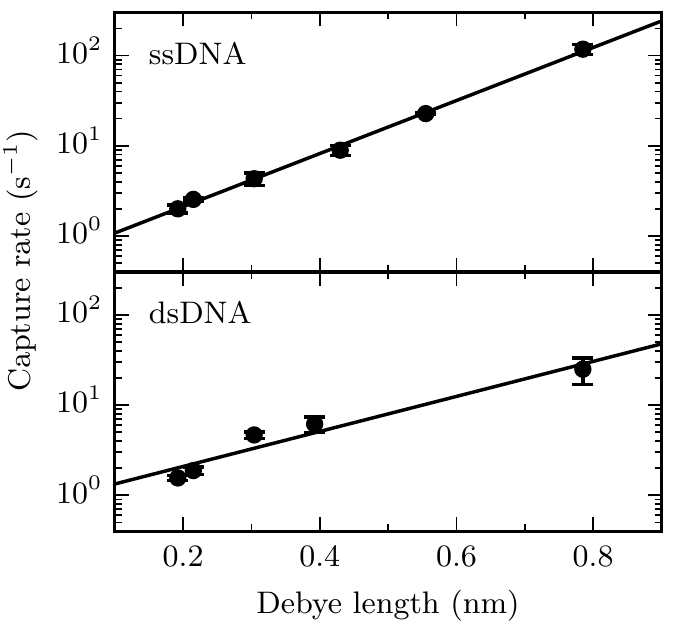}
\caption{Capture rate of ssDNA (top) and dsDNA (bottom) by a ClyA
nanopore at varying ionic strength of the solution 
(data from Ref.~\cite{fran16}).  The data are plotted as a function
of the Debye length~$\ld$, which is related to the salt concentration in
solution $n_0$ as $\ld =\sqrt{\kT/4 \pi e^2 n_0}$.  In the experiments the
NaCl concentration was varied in the range $[0.15,2.5]$~M, corresponding
to a range $[0.2,0.8]$~nm for the Debye length. The solid lines
were obtained from least-squares fitting of an exponential function
$k_\mathrm{on} \sim \exp(\ld/s)$, which yielded $s=0.15$~nm (ssDNA)
and $s=0.22$~nm (dsDNA).}
\label{fig:exp}
\end{figure}

Having reviewed the existing theory of DNA capture by nanopore, we will now 
apply it to the experimental data of Ref.~\cite{fran16}. In that study the 
capture rates of both ssDNA and dsDNA by the ClyA nanopore were measured as a 
function of the ionic strength of the solution. In what follows we will show 
that the experimental data in both cases (shown in Fig.~\ref{fig:exp}) are 
not in line with the theory of diffusion-limited capture. Instead, they seem to 
exhibit the exponential dependence predicted by Eq.~\eqref{Seq:kon_ssDNA}, 
suggesting that it is more likely a reaction-limited process. Finally, we will 
show that the fitted exponent is in line, at least in the order of magnitude, 
with the theoretical predictions, further reinforcing our argument.

Since both theories involve the electrophoretic mobility $\mu$ of DNA,
we will first estimate its value. Assuming that DNA is a cylinder of
diameter $b$ with an effective charge per length equal to $-g\alpha
e/a$ (with $a$ the separation between successive phosphate groups),
Ref.~\cite{gros10} estimates the electrophoretic mobility as
\begin{equation}
\mu = \frac{g \alpha e \ld}{\eta\pi a b},
\label{Seq:mu}
\end{equation}
where $\eta=10^{-3}$~kg m$^{-1}$s$^{-1}$ is the water viscosity, and
$\alpha < 1$ is a numerical factor. The latter takes into account that
counterions are bound to the phosphate charges, rendering the effective
charge of DNA smaller than the bare one.  Finally, one should use $g=1$
and $g=2$ for ssDNA and dsDNA, respectively. Since this quantity enters
in both theories, we will calculate $\mu \Delta V$ for both ssDNA and
dsDNA. Using $ab \approx 0.68$~nm and $\Delta V = 70$~mV we obtain
\footnote{For dsDNA it is $a=0.34$~nm and $b=2$~nm, while for ssDNA one
has $a=0.68$~nm and $b\approx1$~nm~\ec{\cite{chi13}}.}
\begin{equation}
\mu\Delta V \approx 5.2g~\ld~\mathrm{m}~\mathrm{s}^{-1},
\label{Seq:dVmu}
\end{equation}
where for simplicity we have taken $\alpha = 1$. Thus, the electrophoretic
mobilities of ssDNA and dsDNA are found to be quite similar.

We will first test whether the data can be described by the theory of 
diffusion-limited capture. Combining Eqs.~\eqref{Seq:defr*} and 
\eqref{Seq:kon} yields $k_\mathrm{on} = 2 \pi \mu \Delta V \widetilde d c_0$, 
so using the experimental concentration $c_0 = 1$~$\mu$M, the characteristic 
length $\widetilde d = 0.25$~nm [see discussion below Eq.~\eqref{Seq:qeff}] and 
Eq.~\eqref{Seq:dVmu} yields
\begin{equation}
 k_\mathrm{on} = 2.5g \times 10^3~\mathrm{s}^{-1},
 \label{Seq:kon_dif}
\end{equation}
where we have used a representative value $\ld = 0.5$~nm for the Debye
length. A comparison with the experimental data of Fig.~\ref{fig:exp}
indicates that this result overestimates the capture rates by two orders
of magnitude.  Note that some of the phosphate DNA charges can be bound
to counterions, leading to $\alpha < 1$. However, to reconcile the data
with diffusion-limited capture, one would need a very small value of
$\alpha \sim 10^{-2}$, which is unlikely. In addition, the data are
not consistent with a linear dependence on $\lambda_\mathrm{D}$, as
expected from a diffusion-limited process ($k_\mathrm{on} \sim \mu \sim
\lambda_\mathrm{D}$).  We, thus, conclude that the capture of both ssDNA
and dsDNA is not diffusion-limited.~\footnote{Ref.~\cite{fran16} suggested
for ssDNA a reaction-limited capture and for dsDNA a diffusion-limited
capture. The latter conclusion was based on an erroneous estimate of
$k_{\rm on}$.} This is in agreement with measurements for dsDNA of
comparable size captured by solid-state nanopores~\cite{wanu10}.

Having excluded a diffusion-limited capture, let us now test 
the other limiting case, that of a reaction-limited process. Combining 
Eqs.~\eqref{Seq:Veff} and \eqref{Seq:kon_ssDNA} one obtains
\begin{equation}
 k_\mathrm{on} \sim \exp \left[
 \frac{\mu}{\widetilde\mu \kT} (V_\mathrm{max} - V_\mathrm{min}),
 \right] 
 \label{Seq:kon_dV}
\end{equation}
where we have defined $V_\mathrm{max} \equiv V(r_\mathrm{max})$ and 
$V_\mathrm{min} \equiv V(r_\mathrm{min})$ the maximum and minimum values
of the electrophoretic potential. In combination with Eq.~\eqref{Seq:mu},
this relation implies an exponential dependence of the capture rate on
the Debye length. This is indeed the observed trend of the experimental
data, as seen in Fig.~\ref{fig:exp}.  For a more quantitative comparison,
we get from Eq.~\eqref{Seq:cisV}
\begin{equation}
V_\mathrm{max} - V_\mathrm{min} = \Delta V \widetilde d 
 \left(\frac{1}{r_\mathrm{max}} - \frac{1}{r_\mathrm{min}}\right)
 \equiv \Delta V \frac{\widetilde d}{\widetilde r},
\label{Vbarr}
\end{equation}
where $\widetilde r$ is a characteristic length, and is expected to be 
comparable to the pore diameter, i.e.\ $\widetilde r \sim d = 6$~nm. Combining 
this with Eq.~\eqref{Seq:kon_dV} yields
\begin{equation}
 k_\mathrm{on} \sim \exp \left(\frac{\mu\Delta V}{\widetilde\mu\kT} 
\frac{\widetilde d}{d} \right).
\label{Seq:kon_almost}
\end{equation}
The only missing element is the determination of the mechanical 
mobility $\widetilde \mu$. For this purpose one may use Stokes' law, which 
gives
\begin{equation}
 \widetilde\mu~\kT = \frac{\kT}{6\pi\eta \rh} \approx \frac{2.2\times 
10^{-19}}{\rh} \mathrm{m}^3\mathrm{s}^{-1},
\label{Seq:mutilde}
\end{equation}
where $\rh$ is the hydrodynamic radius. Combining this with 
Eqs.~\eqref{Seq:dVmu} and \eqref{Seq:kon_almost}, and using once more 
$\widetilde d = 0.25$~nm and $d = 6$~nm, gives $k_\mathrm{on} \sim \exp 
\left(0.98g~\ld \rh~\mathrm{nm}^{-2}\right) \equiv \exp(\ld/s)$, where
\begin{equation}
s = \frac{1~\mathrm{nm}}{0.98g \rh},
\label{Seq:slope}
\end{equation}
is a parameter that can be fitted to the experimental data (see 
Fig.~\ref{fig:exp}). Since the contour length $L = 62$~nm of ssDNA is much 
larger than its persistence length $l_\mathrm p = 2.2$~nm, and if we neglect 
excluded-volume effects, we can approximate it as a sphere of radius
$\sqrt{l_\mathrm p L/3}$ (radius of gyration). Using this for the
estimation of its hydrodynamic radius gives $\rh \approx 7$~nm, from
which we find $s \approx 0.15$~nm. In the case of dsDNA, the contour
length $L = 31$~nm is lower than its persistence length $l_\mathrm
p = 45$~nm, suggesting it behaves more like a rigid rod. If we,
once more, approximate it as a cylinder of diameter $d$, and use the
results of Ref.~\cite{hans04}, we obtain $\rh \approx 5$~nm.  Finally,
plugging this in Eq.~\eqref{Seq:slope} yields $s \approx 0.10$~nm.
These results are in a good agreement with fits of the experimental data
(Fig.~\ref{fig:exp}), which yielded the values $s = 0.15$~nm and $s =
0.22$~nm for ssDNA and dsDNA, respectively, despite the simplicity
of the theory. The agreement further corroborates the validity of the
reaction-limited capture scenario.

\section{Conclusion}

We have reviewed two basic mechanisms of DNA capture by a nanopore:
the diffusion-limited and the reaction-limited capture. The
theoretical description of these mechanisms was developed in
Refs.~\cite{wanu10,gros10,muth10}, and these ideas were tested in
translocation experiments through solid-state nanopores with dsDNA
sequences ranging from 800 to 50,000 base pairs~\cite{wanu10}. The
shortest lengths dsDNA (up to 8,000 base pairs) showed a reaction-limited
capture, characterized by an exponential growth of the capture rate
$k_{\rm on}$ with the sequence length. A second regime, for sequences
longer than 10,000 base pairs, was found to be consistent with a
diffusion-limited capture, in which $k_{\rm on}$ is independent of
the sequence length~\cite{wanu10}. Overall, solid-state nanopore
experiments~\cite{wanu10} were found to be in agreement with the
theoretical framework of dsDNA capture.

Here we tested the theory in a set of experiments with ClyA, a
biological nanopore recently engineered to allow translocation of both
ssDNA and dsDNA at physiological salt concentrations~\cite{fran16}. The
experiments involved short ssDNA and dsDNA sequences ($90$ nucleotides
and base pairs, respectively), and were performed at varying salt
concentration~\cite{fran16}.  Diffusion-limited capture rates estimated
for a nanopore with the ClyA size were shown to be much higher than
experimental measurements for both ssDNA and dsDNA, suggesting for both
a reaction-limited capture (this corrects the erroneous conclusion in
Ref.~\cite{fran16}) Our analysis showed that the experiments are in
quantitative agreement with the theory, which predicts an exponential
dependence $k_{\rm on} \sim \exp(\lambda_{\rm D}/s)$ on the Debye length
$\lambda_{\rm D}$, with the prefactors determined by the local properties
of the barrier. The theoretical estimates for the characteristic
length $s$ for both ssDNA and dsDNA are in agreement with fits to the
experimental data, confirming the validity of the modeling approach. A
consistent picture thus emerges for the capture mechanism of DNA from ClyA
nanopore.  A reaction-limited capture was also found to be in agreement
with the results of Ref.~\cite{wanu10}, and with other studies of ssDNA
capture into $\alpha$HL nanopores~\cite{henr00,mell02,mell03}. Still,
it would be desirable to have more insight on the nature of the
barrier. A question, which could be addressed by additional experiments
or computer simulations of the capture mechanism, similar to those of
Refs.~\cite{fara13,fara14}.

\medskip

{\sl Acknowledgement } -- SN acknowledges financial support from the
Research Funds  Flanders  (FWO  Vlaanderen)  grant  VITO-FWO 11.59.71.7.

\section*{References}

\providecommand{\newblock}{}

\end{document}